\documentclass[aps,prl,twocolumn,showpacs,superscriptaddress,groupedaddress]{revtex4-1}
\usepackage{graphicx}  
\usepackage{dcolumn}   
\usepackage{bm}        
\usepackage{amssymb}   
\usepackage{color}
\begin{document}

\title{Heat fluctuations in an out of equilibrium bath }

\author{J. R. Gomez-Solano}
\email{juan.gomez\_solano@ens-lyon.fr}
\author{A. Petrosyan}%
\author{S. Ciliberto}%
\affiliation{Laboratoire de Physique,  \'Ecole Normale
Sup\'erieure de Lyon, CNRS UMR 5672, 46,
All\'ee d'Italie, 69364 Lyon CEDEX 07, France
}%

\input   
\date{\today}

\begin{abstract}
We measure the energy fluctuations of a Brownian particle confined
by an optical trap in an aging gelatin after a very fast quench
(less than 1 ms). The strong nonequilibrium fluctuations due to
the  assemblage of the gel, are interpreted, within the framework
of fluctuation theorem, as a heat flux from the particle towards
the bath. We derive, from a simple model, an analytical expression
of the heat probability distribution, which fits the experimental
data and satisfies a fluctuation relation similar to that of a
system in contact with two baths at different temperatures.
\end{abstract}

\maketitle
The heat flux between two reservoirs at different temperatures is an important and useful example of an out of
equilibrium process. In small systems this heat flux  is a strongly fluctuating  quantity
and  the probability distribution of these fluctuations  has been
recently widely studied, within the context of fluctuation theorems \cite{gallavotti}.
These studies have been  mainly devoted to the steady state, that
is when the temperatures $T_A$ and $T_B$ of the two reservoirs, $A$
and $B$, are kept constant. In such a case the probability
distribution $P(Q_\tau)$ of exchanging with the reservoir $A$ the
heat $Q_\tau$ in a time $\tau$, is related to the that of
exchanging the quantity $-Q_\tau$ according:
\begin{equation}
\ln {P(Q_\tau)\over P(-Q_\tau)} = \Delta \beta \ Q_\tau
\label{eq_Pq}
\end{equation}
where $\Delta \beta =(1/T_B -1/T_A)/k_B$, $k_B$ is the Boltzmann
constant and $k_B\Delta \beta \ Q_\tau$ can be easily identified as
the entropy production during the time $\tau$
\cite{gallavotti,bodineau,Maes}. This equation
has been derived for several theoretical models \cite{bodineau,Jarzynski_2004} in the stationary regime. However the non stationary case, although very useful for
applications, has been studied only in some specific
models \cite{crisanti}  of systems relaxing towards
equilibrium. Thus one may wonder whether a relation like
Eq.~(\ref{eq_Pq}) may still hold, how it is  eventually modified and
what kind of information on the system  can be
obtained \cite{crisanti}.

These important questions have never been analyzed in any
experiment. Thus the purpose of this letter is to give new insight
to this problem, by measuring the energy fluctuations of a
Brownian particle used as a probe inside a gelatin relaxing
towards its solid-like state (gel), after a very fast quench, from
above to below the gelation temperature $T_{gel}$. The main result
of our investigation is that these fluctuations can be interpreted
as a heat flux from the particle towards the bath. The measured
$P(Q_\tau)$ satisfies an equation formally equivalent to
Eq.~(\ref{eq_Pq}), but in this case $\Delta \beta$ is a decreasing
function of time. The $P(Q_\tau)$ can be fitted by an analytical
expression that we derive from a Langevin equation for a Brownian
particle coupled with an out of equilibrium  bath.
\begin{figure}
\includegraphics[scale=0.5]{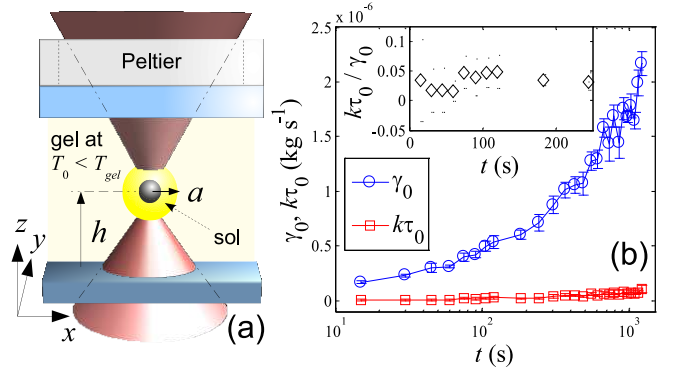}
\caption{\label{fig:1}(a) Schematic representation of the
experimental setup to perform a local quench in a sol droplet
around a trapped particle in the gel bulk. (b) Time evolution of
the viscous drag coefficient  $\gamma_0$ of the particle and the  correlation time $\tau_0$ of the
gelatin droplet measured after the quench at $f=5$ Hz . Inset:
$k\tau_0/\gamma_0$ as a function of time.}
\end{figure}

The experiment has been performed using gelatin,  a
thermoreversible gel obtained from denatured collagen. Above
$T_{gel}$ an aqueous gelatin solution is in a liquid viscous phase
(sol), whereas below $T_{gel}$ the formation of a network of
cross-linked filaments leads to an elastic solid-like phase (gel)
\cite{djabourov}. In this gel phase the gelatin viscoelastic
properties slowly evolve toward equilibrium and share some common
phenomenological features with glassy dynamics
\cite{parker,ronsin,normand}.  We are interested in this transient
out-of-equilibrium regime, that we use to  study the fluctuations
of the energy fluxes from and to the heat bath in the
nonstationary case. A similar problem has been theoretically
analyzed for the first time in Ref.~\cite{crisanti} for a
model of aging spin glasses. It has been found that a relation
like Eq.~(\ref{eq_Pq}) can be applied to a relaxing system to
obtain quantitative informations on the heat exchanges with the
bath. We show in this letter that this approach, exploited only
once in real experiments \cite{jop}, can be indeed very useful for
understanding the properties of a Brownian particle in an
out-of-equilibrium bath.
\begin{figure}
\includegraphics[scale=0.625]{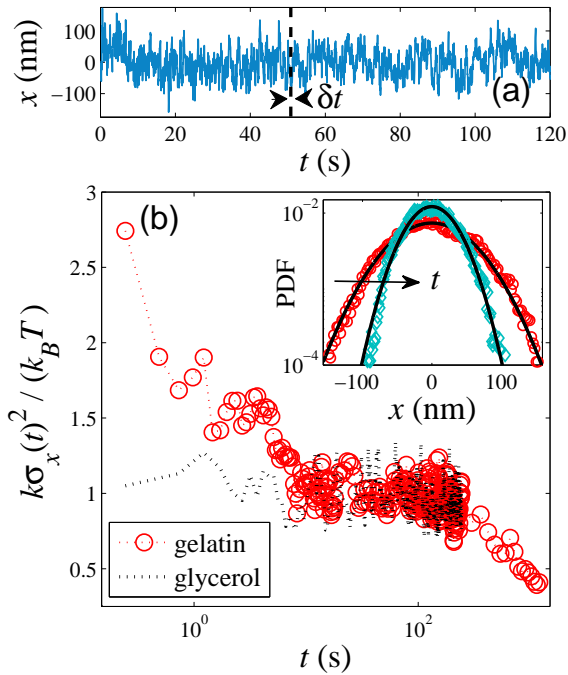}
\caption{\label{fig:2}(a) Time evolution of $x$ after a quench.
$\sigma_x(t)^2$ is computed over $\delta t=0.1$ s and over 60
independent quenches. (b) Time evolution of $\sigma_x(t)^2$
(normalized by $k_B T/k$) after the quenches performed in gelatin
($\circ$) and glycerol (dashed line). Inset: Probability density
of $x$ at $t=$ 0.5 s and 50 s for the quench in gelatin. The solid
lines are Gaussian fits.}
\end{figure}

In the present experiment, an aqueous gelatin solution (type-B pig
skin) at a concentration of 10\%wt is prepared following the usual
protocol \cite{normand}. For this sample $T_{gel} = 29^{\circ}$C.
This solution fills a transparent cell the temperature of which is
controlled by a Peltier element [Fig.~\ref{fig:1}(a)], at
$T_0=26\pm 0.05^\circ$C $<T_{gel}$. Thus the solution inside the
cell is in the solid-like phase.  A silica bead, of radius
$r=1\,\mu$m, is kept inside the gelatin in the focal position of a
tightly focused laser beam ($\lambda=980$ nm) at a power of 20 mW.
At this power the laser produces on the particle an elastic force
of stiffness $k=2.9 \ \rm{pN}/\mu$m. Because of light absorption,
the temperature of the trapped particle is $T=27^\circ$C, which is
still smaller than $T_{gel}$. Therefore the bead  is inside the
solid gel in the beam focus at a distance $h=25\,\mu$m from the
cell wall, see Fig.\ref{fig:1}(a) \cite{experiment}. Starting from
this condition, the laser power is increased to 200 mW and the
local temperature around the focus rises to $38^{\circ}$C
$>T_{gel}$. As a result the gel melts and a liquid droplet of
radius $a=5\,\mu$m,
 is formed around the trapped bead inside the
the solid gel bulk, as sketched in Fig.~\ref{fig:1}(a). After
$180$ s, the laser power is suddenly decreased again to 20 mW so
that the temperature is homogenized by heat diffusion into the
bulk in less than 1 ms resulting in a very efficient quench of the
droplet to the  final  temperature $T<T_{gel}$. At $T$
the liquid inside the droplet  solidifies in about 1 hour and the
particle, trapped in the center of the drop by the focused beam,
is a probe of this relaxation dynamics. The quenching procedure is
repeated 60 times in order to perform the proper ensemble
averages.

Immediately after the quench we record the time evolution of the
$x$ position [see Fig.~\ref{fig:1}(a)] of the trapped particle
measured by a position sensitive detector whose output is sampled
at 8 kHz and acquired by a computer. The resolution of the
measurement of $x$ is about $1$ nm \cite{resolution,jop}. In order to
characterize the particle dynamics we measure,  using active microrheology \cite{jop}, the time evolution
of the viscous drag coefficient $\gamma_0$ of the particle and the largest correlation  time $\tau_0$ of
the fluid. This is done by measuring the response of the bead at a
time-dependent sinusoidal force $F$ of amplitude $87 \,\rm{fN}$ and
frequency $f$ applied to the bead. The force $F=k  x_0$ is
obtained through the modulation  of the beam focus position $x_0$.
The results for $\gamma_0$ and $k\tau_0$, measured at $f=5$ Hz, are
shown in Fig.~\ref{fig:1}(b). First, for $t \lesssim 200$ s after the quench
there is a transient regime where the droplet is purely viscous,
$\tau_0\simeq 0$, whereas  $\gamma_0$ increases in time. In this
regime $\gamma_0$ and $\tau_0$ do not depend on $f$.  For $t>200$
s the liquid gelatine inside the drop has a behavior similar to
that observed in macroscopic samples \cite{normand,djabourov}, \emph{i.e.}
the liquid drop is actually undergoing gelation. We
will study the nonequilibrium statistical properties of the bead
dynamics in the very first $200$ s after the quench where the
liquid gelatin inside the drop is mainly viscous and the
elasticity   is negligible with respect to $k$, as shown in the
inset of \ref{fig:1}(b), where we plot $k\tau_0/\gamma_0$ as a
function of time.

We begin by analyzing  the variance $\sigma_x(t)^2$ of $x$ at time
$t$ after the quench. $\sigma_x(t)^2$ is computed over 60
independent quenches and over a short time window $\delta t = 0.1$
s around each value of $t$ in order to improve the statistics, as
depicted in Fig.~\ref{fig:2}(a). The time evolution of
$\sigma_x(t)^2$ is plotted in Fig.~\ref{fig:2}(b). At the
beginning, $\sigma_x(t)^2$ is almost three times the equipartition
value $k_B T/k$ that would be obtained  at equilibrium. This shows
the presence of a  stochastic force on the particle  due to the
transient formation of the gel network. This force weakens
compared to the thermal fluctuations becoming negligible at $\sim
20$ s so that $\sigma_x(t)^2$ slowly decreases in time, reaching
the equilibrium value for $t\gtrsim 20$ s. This relaxation timescale is
two orders of magnitude larger than the initial viscous relaxation
time of the particle: $\tau_k = \gamma_0/ k = 65$ ms. Finally for
$t \gtrsim 200$ s,  $\sigma_x(t)^2$ starts again to decrease
because of  the appearance of a strong elastic component of the
gel confirming the direct measure of $\gamma_0, \tau_0$, shown in
Fig.~\ref{fig:1}(b),  and justifying that for $t\le 200$ s the
gelatin  elasticity is negligible. During this relaxation
process $x$ remains Gaussian as shown in the inset of
Fig.~\ref{fig:2}(b).

In Fig.~2(b) we also plot the time evolution of $\sigma_x(t)^2$
measured, after the same quenching procedure, in a Newtonian fluid
(glycerol 60\%wt in water) with the same viscosity of the initial
sol phase of gelatin. In this case, the particle dynamics
must settle into an equilibrium state in a time $\sim \tau_k $
after the quench \cite{chetrite}. Indeed in Fig.~\ref{fig:2}(b) we
see that, in glycerol,  $\sigma_x(t)^2 = k_B T / k$ for all $t$
within the experimental accuracy. This confirms that no
experimental artifact is present and that the observed dependence
of $\sigma_x(t)^2$ in gelatin  is a real non-equilibrium effect
due to the sol-gel transition.

\begin{figure}
\includegraphics[scale=0.65]{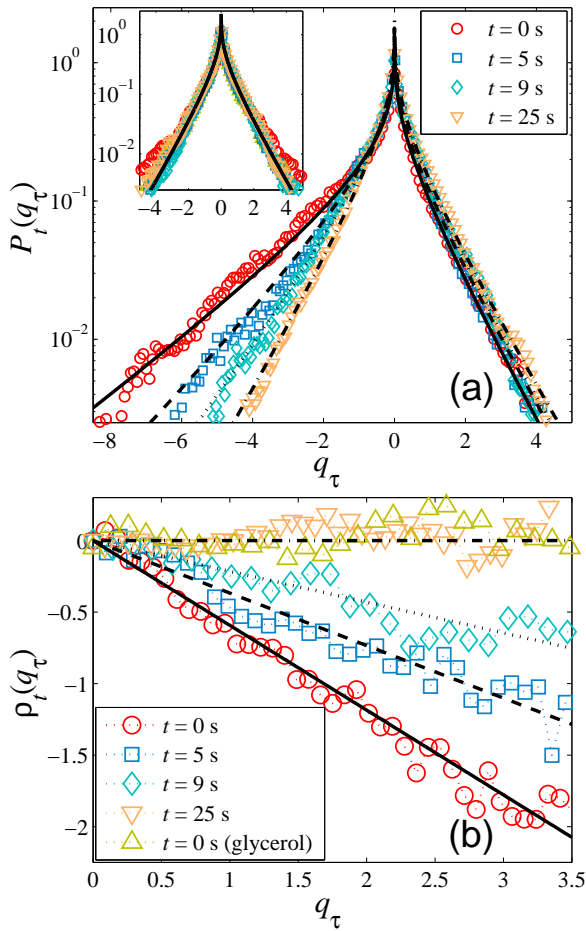}
\caption{\label{fig:3} (a) Probability density $P_t(q_{\tau})$ of
the normalized heat $q_{t,\tau}$ for $\tau=30$ s at different
times $t$ after the quench. The theoretical lines are computed
using Eq.~(\ref{eq:PDFheat}). Inset:  $P_t(q_{\tau})$ computed, at
the same $t$ and $\tau$  during a quench performed in glycerol.
The solid line corresponds to the theoretical equilibrium profile.
(b) Asymmetry function of the $P_t(q_{\tau})$ shown in (a). The
straight lines are obtained using Eq.~(\ref{eq:symmetry1}).}
\end{figure}

We now focus on the energy fluctuations of the particle inside the
droplet  for $t<200$ s, \emph{i.e.} when  $k\tau_0/\gamma_0 \ll 1$. Thus the
energy of the particle is simply $U(t)=kx(t)^2/2$. As there is
no external force acting to the particle,  the energy fluctuation
$\Delta U_{t,\tau}$  during the time $\tau$ is equal to the heat
$Q_{t,\tau}$  exchanged between  the particle and the bath
\cite{sekimoto}, specifically
\begin{equation}\label{eq:balance}
Q_{t,\tau} = \Delta U_{t,\tau} =\frac{k}{2}(x_{t+\tau}^2-x_t^2).
\end{equation}
The mean heat transferred during $[t,t+\tau]$ is \\ $\langle
Q_{t,\tau} \rangle = (k/2)[\sigma_x(t+\tau)^2-\sigma_x(t)^2] \le
0$,
 which reveals the
existence of a mean heat flux from the particle to the
surroundings over the timescale $\tau$ because of the relaxation
of $\sigma_x$. The maximum value $|\langle Q_{t,\tau} \rangle |
\approx k_B T$ takes place at $t=0$ s and for 20 s $\lesssim \tau
\lesssim$ 200 s. Non-negligible values of the mean heat compared to
$k_B T$ persist for several seconds after the quench.
Nevertheless, as $t$ increases, $|\langle Q_{t,\tau} \rangle |$
decreases becoming negligible and experimentally undetectable for
$t \gtrsim$ 20 s .

The probability density function $P_t(q_{\tau})$ of the normalized
heat $q_{t,\tau} = Q_{t,\tau}/(k_B T)$
is computed over the 60
 quenches and over a short time window $\delta t = 0.1$ s
around each $t$ and $t+\tau$, as sketched in Fig~\ref{fig:2}(a).
We focus on a large value of $\tau$ in order to probe timescales
comparable to the relaxation of the nonthermal fluctuations.
Fig.~\ref{fig:3}(a) shows  $P_t(q_{\tau})$ at different times $t$
after the quench for $\tau=30$~s. $P_t(q_{\tau})$ is highly
non-Gaussian with a spike at $q_{\tau}=0$ and slowly decaying
tails for all the values of $t$. Immediately after the quench,
$P_t(q_{\tau})$ is strongly asymmetric with a long tail occurring
at negative fluctuations. As $t$ increases this  asymmetry
decreases and $P_t(q_{\tau})$ becomes  symmetric at $t \gtrsim 20$
s.Once again, we check that the long-lived asymmetry occurs
because of the intricate nonequilibrium nature of the bath. In the
inset of Fig.~\ref{fig:3}(a) we plot  $P_t(q_{\tau})$ with
$\tau=30$ s for the local quenches performed in glycerol.
$P_t(q_{\tau})$ quickly converges to the equilibrium  profile and
it is always symmetric with respect to $q_{\tau}=0$.

As in Eq.(\ref{eq_Pq}), the \emph{asymmetry function}
$\rho_t(q_{\tau})=\ln\left[{P_t(q_{\tau})}/{P_t(-q_{\tau})}\right)]$
is  commonly used to measure the asymmetry of $P(q_\tau)$ between
the positive and the negative values of the fluctuations. The
function $\rho_t(q_{\tau})$, computed from the $P_t(q_{\tau})$
shown in Fig.~\ref{fig:3}(a), is plotted in Fig~\ref{fig:3}(b). It
is a linear function of its argument $q_{\tau}$: $\rho_t(q_{\tau})
= - \Delta\beta_{t,\tau} q_{\tau}$. The slope
$\Delta\beta_{t,\tau}$ decreases as $t$ increases approaching the
symmetric value $\Delta\beta_{t,\tau} = 0$ as $|\langle Q_{t,\tau}
\rangle| \ll k_B T$. This linear relation, except for the time
dependent $\Delta \beta_{t,\tau}$, is formally similar to
Eq.(\ref{eq_Pq}), and it is the first experimental evidence of the
phenomenon  theoretically obtained  for a relaxing spin glasses in
Ref.~\cite{crisanti}. For comparison we also plot
$\rho_t(q_{\tau})$ for the quench in glycerol at $t=0$ and
$\tau=30$ s. In this case the heat exchange process is always
symmetric, stressing that this is due to
the nonequilibrium nature of the bath.

In absence of a  theory for our experimental results we model the
nonequilibrium dynamics of the particle  by an  overdamped Langevin
equation  for $x$:
\begin{equation}\label{eq:Langevin}
 \gamma_0 \ \dot{x}_t  =-kx_t + \zeta_t, \ \  \ \  \ \  {\rm{for}} \ t <200\,\rm{s}
\end{equation}
where $-kx_t$ is the harmonic force exerted by the optical trap,
and  $\zeta_t$  is a random noise representing the interaction of
the particle with the out of equilibrium bath, \emph{i.e.} the drop
undergoing gelation. Because of the out-of-equilibrium state,  the
statistical properties of $\zeta$ are unknown, and  it is in
general a nonstationary and correlated process.
 Multiplying Eq.~(\ref{eq:Langevin}) by
$\dot{x}_t$ and integrating over the time interval $[t,t+\tau]$
one obtains the energy balance Eq.~(\ref{eq:balance}) \cite{sekimoto}, where $Q_{t,\tau}$  is
\begin{equation}\label{eq:heat}
Q_{t,\tau} = -\int_t^{t+\tau} \gamma_0 \  \dot{x}_{s}^2   \ \mathrm{d}s
 + \int_t^{t+\tau} \zeta \dot{x}_{s}\, \mathrm{d}s,
\end{equation}
which is the sum of the viscous  dissipation plus the the heat
injected by the bath. Notice that $Q_{t,\tau}$ cannot be estimated
directly as $\zeta$ is unknown. The only experimental way to
measure $Q_{t,\tau}$ is via Eq.~(\ref{eq:balance}).

The asymmetry of $P_t(q_{\tau})$ can be directly linked to the
nonstationarity of the aging bath through the quantity $\sigma_x$.
Using Eq.~(\ref{eq:balance}) and the experimental fact that $x$ is
Gaussian [Fig.~\ref{fig:2}(b)], the analytical expression of
$P_t(q_{\tau})$ for large $\tau$ can be computed \cite{experiment}:
\begin{equation}\label{eq:PDFheat}
P_t(q_{\tau})=\frac{A_{t,\tau}}{\pi}K_0\left(B_{t,\tau}|q_{\tau}|\right)
\exp\left(-\frac{\Delta_{t,\tau} A_{t,\tau}}{2}  q_{\tau}\right),
\end{equation}
where $K_0$ is the zeroth-order modified Bessel function of the
second kind,
\begin{eqnarray}\label{eq:A}
\Delta_{t,\tau}=\frac{\sigma_x(t)}{\sigma_x(t+\tau)}-\frac{\sigma_x(t+\tau)}{\sigma_x(t)},
 \ \ A_{t,\tau}=\frac{k_BT}{k\sigma_x(t)\sigma_x(t+\tau)} \nonumber
\\
{\rm{and}} \ \ \ B_{t,\tau}=A_{t,\tau}\sqrt{1+{\Delta_{t,\tau}^2}/4}
\ \ \ \ \ \ \ \ \ \ \ \ \ \ \ \ \ \ \nonumber
\end{eqnarray}
In Eq.~(\ref{eq:PDFheat}) the asymmetry of the density is
completely determined by the parameter $\Delta_{t,\tau}$ in the
exponential. At equilibrium $\Delta_{t,\tau} = 0$,
$A_{t,\tau}=B_{t,\tau}=1$ regardless of $t$ and $\tau$, so that
one recovers the symmetric equilibrium profile
$P_t(q_{\tau})=K_0(|q_{\tau}|)/\pi$ with $\langle q_{t,\tau}
\rangle = 0$ \cite{imparato}. In Fig.~\ref{fig:3}(a)
for each experimental $P_t(q_{\tau})$ we plot the theoretical
prediction given by the analytical formula (\ref{eq:PDFheat})
using the respective experimental values of $\sigma_x$ shown in
Fig.~\ref{fig:2}(b). The excellent agreement confirms that the
Langevin model (\ref{eq:Langevin}) is suitable to describe the
particle dynamics and the heat exchange with the
gelatin bath after the quench.

From Eq.~(\ref{eq:PDFheat}) one obtains the explicit expression
for the asymmetry function
$\rho_t(q_{\tau})=-\Delta\beta_{t,\tau}q_{\tau}$ and
$\Delta\beta_{t,\tau}$
\begin{equation}\label{eq:symmetry1}
\Delta\beta_{t,\tau}=
\frac{k_BT}{k}\left[\frac{1}{\sigma_x(t+\tau)^2}-\frac{1}{\sigma_x(t)^2}\right].
\end{equation}
Hence, the linearity of $\rho_t(q_{\tau})$ is analytically
satisfied for all the values of the heat fluctuations and for all
$t$ even when $P(q_\tau)$ is strongly non-Gaussian. In
Fig.~\ref{fig:3}(b) we plot the straight lines  with the slope
$\Delta \beta_{t,\tau}$ given by Eq.~(\ref{eq:symmetry1}) and
computed using the experimental values of $\sigma_x$. The good
agreement  with the experimental data shows that
Eq.~(\ref{eq:PDFheat}) verifies  a fluctuation relation, as
Eq.~(\ref{eq_Pq}).

Eq.(\ref{eq:symmetry1}) gains a very intuitive interpretation if
one introduces an equipartition-like relation for the particle
motion for $0\le t\lesssim 60$ s: $k_B T_{eff}(t)=k\sigma_x(t)^2$.
Here $T_{eff}$ is the effective temperature perceived by the
particle due to its coupling with the nonequilibrium gelatin
environment. In this way the parameter $\Delta\beta_{t,\tau}$ can
be written conveniently as
$\Delta\beta_{t,\tau}=\left[ 1/T_{eff}(t+\tau)-1/T_{eff}(t)
\right]T,$
which is formally equivalent to that of Eq.~(\ref{eq_Pq}). Hence
$\Delta S_\tau=-k_B \Delta\beta_{t,\tau} q_{t,\tau}$ can be
naturally identified as the entropy produced by the breakdown of
the time-reversal symmetry due to the the effective temperature
imbalance at two different times after the quench. As the gelatin
droplet ages $\Delta S_\tau$ slows down and the particle exhibits
an equilibrium-like dynamics for the experimental timescales. We
point out that unlike Eq.~(\ref{eq_Pq}) derived in
Refs.~\cite{bodineau,Jarzynski_2004} for
nonequilibrium steady states, Eq.~(\ref{eq:symmetry1}) holds for a
nonstationary regime created by the nonequilibrium bath.

In conclusion, we have experimentally studied the fluctuations of
the heat exchanged between a trapped Brownian particle and a
non-stationary bath, i.e. an aging gelatin  after a very fast
quench.
We have shown that the distribution of the heat satisfies a
fluctuation relation even when the bath is in a non-stationary
state. A Langevin model justifies the observation. The analogy of
our results with those obtained for spin glasses  suggests that
this fluctuation relation may appear as a very robust symmetry
property of  heat exchange processes in other kinds of relaxing
systems.


\begin{thebibliography}{99}
\bibitem{experiment} more details about the experiment and the derivation of
Eq.~(\ref{eq:PDFheat}) will be
given in a longer report.
\end{thebibliography}
\end{document}